\documentclass[11pt]{article}
\usepackage{amsfonts,amsmath,amssymb}
\usepackage{color}
\usepackage{epic}
\usepackage{graphics}
\usepackage{boxedminipage}
\usepackage{epsfig}
\usepackage{changebar}
\usepackage{lscape}

\numberwithin{equation}{section}

\newcommand{\invdots}{{\mathinner{\mkern1mu\raise1pt
   \vbox{\kern7pt\hbox{.}}\mkern2mu\raise4pt\hbox{.}
   \mkern2mu\raise7pt\hbox{.}\mkern1mu}}}

\newtheorem{definition}{Definition}[section]

\newtheorem{proposition}[definition]{Proposition}

\newtheorem{corollary}[definition]{Corollary}

\definecolor{dcyan}{rgb}{0,.8,.8}
\definecolor{ddcyan}{rgb}{0,.6,.6}
\definecolor{dgreen}{rgb}{0,.8,0}
\definecolor{lyellow}{cmyk}{0,0,.2,0}
\definecolor{rose}{rgb}{1,.6,.6}
\definecolor{violet}{rgb}{.9,0,.6}

\newcommand{\1}{\mbox{\hspace{.0em}1\hspace{-.2em}I}}

\def\RR{{\mathbb R}}
\def\ZZ{{\mathbb Z}}
          \def\cB{{\cal B}}          
          \def\cE{{\cal E}}          
          \def\cH{{\cal H}}          \def\cI{{\cal I}}
                    \def\cL{{\cal L}}
\def\cM{{\cal M}}                    
          \def\cQ{{\cal Q}}          
          \def\cT{{\cal T}}          \def\cU{{\cal U}}
\def\cV{{\cal V}}                    
\def\cY{{\cal Y}}          

\newcommand{\lda}{\lambda}

\newcommand{\ie}{{\it i.e.}\ }

\def\tphi{{\widetilde{\phi}}}
\def\tPi{{\widetilde{\Pi}}}
\def\tpi{{\widetilde{\pi}}}

\def\fa{{\mathfrak a}}

\newcommand{\be}{\begin{eqnarray}}
\newcommand{\ee}{\end{eqnarray}}
\newcommand{\bea}{\begin{eqnarray}}
\newcommand{\eea}{\end{eqnarray}}

\begin{document}

\vspace{8mm}

\begin{center}
   {\Huge  {\sffamily Multisymplectic approach to integrable defects in the sine-Gordon model
  }}\\[1cm]

\vspace{8mm}

{\large \textbf{Vincent Caudrelier}}\\
\vspace{1cm}
Department of Mathematics,\\
City University London, \\
Northampton Square, EC1V 0HB London

\end{center}

\vspace{1cm}

 \begin{abstract} 
 Ideas from the theory of multisymplectic systems, introduced recently in integrable systems by the author and Kundu to discuss 
 Liouville integrability in classical field theories with a defect, are applied to the sine-Gordon model. The key ingredient 
 is the introduction of a second Poisson bracket in the theory that allows for a Hamiltonian description of the model that is 
 completely equivalent to the standard one, in the absence of a defect. In the presence of a defect described by frozen 
 B\"acklund transformations, our approach based on the new bracket unifies the various tools used so far to attack the problem. It also gets rid of 
 the known issues related to the evaluation of the Poisson brackets of the defect matrix which involve fields at coinciding space point 
 (the location of the defect). The original Lagrangian approach also finds a nice reinterpretation in terms of 
 the canonical transformation representing the defect conditions.  
  \end{abstract}

\vspace{1cm}



\pagestyle{plain}
\setcounter{footnote}{0}
\section{Introduction}

Integrable systems are a privileged area of Physics and Mathematics where one can test ideas on "toy models" that are sufficiently simple to 
be amenable to exact analytic treatment but sufficiently complex to capture interesting physical phenomena. When studying a model, the 
question of defects or impurities is an important (and often difficult) one, for at least two reasons: 
they represent the departure from an ideal system towards a more realistic situation and they can have dramatic effects on the predicted 
ideal behaviour. Therefore, there is a strong motivation to study defects/impurities in the context of integrable systems. 

Initially, the focus was mainly on quantum (field) theories \cite{DMS}-\cite{CMR} and
remained concentrated until quite 
recently \cite{BG}-\cite{W} on various quantum systems. The point of view taken in those works was to \textit{maintain} integrability
in the presence of defects. The general framework that includes
the previous studies was  proposed in \cite{CMRS}.

 The question of integrable defects in classical field
theories was initiated in a series of papers \cite{BCZ}-\cite{CZ} related to several key
models like the sine-Gordon model, the nonlinear Schr\"odinger (NLS) equation, etc. 
Within a Lagrangian formulation, the idea was to require that the contribution from the defect compensates
for the loss of conservation of the momentum due to its presence. It was argued
that this is enough to ensure the  integrability of a defect model. 
  The defect boundary conditions on the
fields obtained in this way were recognized as B\"acklund transformations
frozen at the location of the defect. This approach triggered a strong
activity in the analysis of the defect in integrable classical field
theories. The observation on frozen B\"acklund transformations was fully exploited in
\cite{VC} in conjunction with the Lax pair formulation of the general AKNS
approach \cite{AKNS} to obtain a generating function of the entire set of
modified conserved quantities. This settled the question of integrability in the sense of having 
an infinite number of local conserved charges. The approach also allowed to answer some questions
left open in the Lagrangian formulation, like the formulation of the defect
conditions directly in terms of the fields of the theory for models like KdV. 
The question of Liouville integrability, however, remained opened. The direct use of Poisson brackets
to implement the method of the classical $r$-matrix is hindered by divergences due to the localized defect. 
Some regularization is needed. The sine-Gordon model was first
studied in \cite{Hab-kundu}. Later on, a very nice
series of papers tackled the question systematically for several models
\cite{avanDaik13,anvanSG10,Doiku-NLS}. The procedure is based on the a priori assumption that the defect matrix
satisfies appropriate Poisson bracket relations. A careful regularization then yields the so-called "sewing conditions" between
the fields in the bulk and those contained in the defect matrix. The
consistency of the approach must then be checked a posteriori.

At that stage, there were essentially two approaches to the same question that were not 
quite reconciled: the Lagrangian/B\"acklund approach and the "sewing conditions" approach.
This situation was the motivation for the introduction in \cite{CK} of the multisymplectic formalism 
to study Liouville integrability of classical field theories with a defect. This was done 
in detail for the case of the NLS equation. The purpose of the present paper is 
to apply these ideas to another famous prototype of integrable field theory: the sine-Gordon model. 
The main motivation is to test how general the ideas developed for NLS in \cite{CK} are. The sine-Gordon model is 
widely recognized as one of the most important representatives of the class of relativistic integrable field theories, while 
the NLS equation has a similar status in the class of non-relativistic integrable models. 
It is shown that they apply just as successfully, providing the same unifying picture as for the 
NLS case and getting rid of the known open problems. 

The paper is organised as follows. In Section \ref{multisymp}, the two Poisson brackets at the basis
 the multisymplectic formalism are introduced for the sine-Gordon model and it is shown that two completely equivalent 
Hamiltonian descriptions exist for this model. Conservation laws and the classical $r$ matrix approach are discussed in the 
light of the two Poisson brackets, each one corresponding to an independent variable (called space and time variable). In Section 
\ref{defField}, we review the sine-Gordon model with a defect from the Lagrangian and Lax pair points of view. 
We then go on to show how the new formalism allows to prove Liouville integrability of the model with defect directly, with no 
resort to sewing conditions. The key point is the fact that the defect conditions are reinterpreted as a canonical transformation with respect to the 
new Poisson structure. It turns out that the generating functional for this canonical transformation is directly built on the defect Lagrangian 
density originally derived to study problems with defects. The new classical $r$ matrix approach with defect is also derived there and the new 
monodromy matrix containing the generating function for the conserved quantities in involution is exhibited.
Conclusions and outlooks are gathered in Section \ref{conclusions}

\section{Multisymplectic structure of the sine-Gordon model}\label{multisymp}

\subsection{Poisson brackets and Hamiltonian equations}\label{multi_sG}

Let us present the multisymplectic structure of the sine-Gordon model by introducing two Poisson brackets on the
phase space of the model. The main observation behind the multisymplectic
approach to field theory is that the canonical quantization procedure puts
emphasis only on the time parameter and, as a consequence, only considers a
partial Legendre transformation when defining canonical conjugate
coordinates. The idea of the multisymplectic approach is to restore the balance between the independent variables and 
to consider one Legendre transformation per independent variable. Although this research area has developed in a rather non systematic 
 way (see e.g. \cite{BH} for an attempt to give an account of the various
approaches) and into a heavy mathematical formalism, the commonly accepted
origin is the so-called De Donder-Weyl formalism \cite{DDW}.

In the traditional approach, given fields
$\phi_a$ depending on the independent variables $(x,t)$\footnote{It is enough for our purposes to consider $1+1$
dimensional field theory.} one defines the conjugate momenta $\pi^a$ as 
\bea
\label{Legendre1} \pi^a=\frac{\partial \cL}{\partial (\partial_t \phi_a)}\,,
\eea 
where $\cL$ is the Lagrangian density of the theory. Then, one imposes
\textit{equal-time} canonical relations by defining the {\it space} 
Poisson brackets as
\bea 
\{\phi_a(x,t_0),\pi^b(y,t_0)\}_S=\delta_a^b\delta(x-y)~~,~~\{\phi_a(x,t_0),\phi_b(y,t_0)\}_S=0~~,~~\{\pi^a(x,t_0),\pi^b(y,t_0)\}_S=0\,, 
\eea 
at some initial time $t_0$.  The subcript $S$ on
the Poisson bracket indicates that it is equal-time \ie it does not depend
on time but only on space.  However, the Legendre transformation
(\ref{Legendre1}) is incomplete and one can also define another set of
conjugate momenta by 
\bea 
\label{Legendre2} 
\Pi^a=\frac{\partial
\cL}{\partial (\partial_x \phi_a)}\,.  
\eea 
The second bracket is then defined in complete analogy by 
\bea
\{\phi_a(x_0,t),\Pi^b(x_0,\tau)\}_T=\delta_a^b\delta(t-\tau)~~,~~\{\phi_a(x_0,t),\phi_b(x_0,\tau)\}_T=0~~,~~\{\Pi^a(x_0,t),\Pi^b(x_0,\tau)\}_T=0\,, 
\eea at some
fixed "initial" location $x_0$. 
These relations can
be seen as {\it equal-space canonical brackets}.  The subscript $T$
indicates that this Poisson bracket does not involve space.  These two
brackets form the basis of the formulation of covariant
Poisson brackets for field theories.

We now develop this formalism for the sine-Gordon model. We show that the two Hamiltonian pictures corresponding to the two brackets 
 $\{~,~\}_{S}$ and  $\{~,~\}_{T}$ yield completely equivalent descriptions of the sine-Gordon model. 
It is very important to note that the multisymplectic formalism
presented here is very different from the well-known bi-Hamiltonian theory of
integrable systems \cite{Magri}. The latter is based on
the existence of two compatible { equal-time } brackets $\{~,~\}_{S1}$ and
$\{~,~\}_{S2}$, each of which allows for the description of the {\it time}
evolution of the model.  Our equal-space bracket  $\{~,~\}_{T}$ 
on the other hand yields the {\it space}
evolution of the model. 

The sine-Gordon model (in laboratory coordinates) is a relativistic field theory for the scalar field $\phi(x,t)$ with equation of motion given by
 \bea 
 \label{sG}
 \phi_{tt}-\phi_{xx}+\frac{m^2}{\beta}\sin \beta\phi=0\,, 
\eea 
where $m$ is the mass parameter and $\beta$ the coupling constant. 
A Lagrangian density for this equation is 
\bea
\cL=\frac{1}{2}(\phi^2_{t}-\phi_{x}^2)-\frac{m^2}{\beta^2}(1-\cos\beta\phi)\,. 
\eea 
From this, applying the Legendre transformations discussed above, we get the conjugate momenta as
\bea
\pi=\frac{\partial \cL}{\partial \phi_t}=\phi_t~~,~~\Pi=\frac{\partial \cL}{\partial \phi_x}=-\phi_x\,.
\eea 
The associated Hamiltonian densities then read
\bea
\cH_S&=&\frac{1}{2}\pi^2+\frac{1}{2}\phi_x^2+\frac{m^2}{\beta^2}(1-\cos\beta\phi)\,,\\
\label{Ham_T_density}
\cH_T&=&-\frac{1}{2}\Pi^2-\frac{1}{2}\phi_t^2+\frac{m^2}{\beta^2}(1-\cos\beta\phi)\,.
\eea
Given the canonical Poisson brackets
\bea 
\label{space_brackets}
\{\phi(x,t_0),\pi(y,t_0)\}_S=\delta(x-y)~~,~~\{\phi(x,t_0),\phi(y,t_0)\}_S=0~~,~~\{\pi(x,t_0),\pi(y,t_0)\}_S=0\,, 
\eea
and 
\bea
\label{time_brackets}
\{\phi(x_0,t),\Pi(x_0,\tau)\}_T=\delta(t-\tau)~~,~~\{\phi(x_0,t),\phi(x_0,\tau)\}_T=0~~,~~\{\Pi(x_0,t),\Pi(x_0,\tau)\}_T=0\,, 
\eea
one easily checks that \eqref{sG} is recovered from the following Hamiltonian equations of motion
\be
\label{time_eqs_motion}
\begin{cases}
\phi_t=\{\phi,H_S\}_S\,,\\
\pi_t=\{\pi,H_S\}_S\,,
\end{cases}
\ee
or
\be
\label{space_eqs_motion}
\begin{cases}
\phi_x=\{\phi,H_T\}_T\,,\\
\Pi_x=\{\Pi,H_T\}_T\,,
\end{cases}
\ee
where
\be
\label{Hams}
H_S=\int\cH_S\,dx~~,~~H_T=\int\cH_T\,dt\,.
\ee
\underline{Important remark:} Given the natural symmetry of space and time coordindates for the sine-Gordon model, the reader may have the impression that 
the above discussion (and some of the results to come concerning the Lax pair formulation, integrals of motion and classical $r$ matrix approach)
is just a trivial exercise in rewriting the theory under the swap $x\leftrightarrow t$ and the change $\phi\to -\phi$. Our point is that, in the 
traditional approach, even after such a transformation, one would still use a single Poisson structure describing time evolution of the 
model. In the present approach, after this transformation, the key point is that the two Poisson brackets still coexist 
and bring complementary information on the time and space evolution of the system. It is the (co)existence of these Poisson brackets, irrespective 
of the chosen coordinates, that is the crucial ingredient of the multisymplectic approach. 
Moreover, in the presence of a defect, no matter what coordinate is declared as the space coordinate, 
once this is fixed, the traditional approach based on a single Poisson bracket still has the limitations explained in the introduction. In contrast, 
the present approach is applicable and provides the missing tool, \ie the new equal-space Poisson bracket, 
to tackle the question of Liouville integrability, as we proceed to explain in the rest of this paper.

\subsection{Conservation laws and space and time integrals of motion}\label{CL}

Parallel to the traditional Hamiltonian formalism for sine-Gordon, it is well-known that this model is integrable in the sense that it can be formulated 
via a Lax pair and the associated zero-curvature representation. Following for instance \cite{FT}, we consider the auxiliary problem 
\bea
\label{aux_pb}
\begin{cases}
\Psi_x=U\,\Psi\,,\\
\Psi_t=V\,\Psi
\end{cases}
\eea
where $U$ and $V$ are two $2\times 2$ matrices depending on $x$, $t$ and the so-called spectral parameter $\lda$, given by
\bea
\label{LP1}
U(x,t,\lda)&=&-i\frac{\beta}{4}\pi\,\sigma_3-ik_0(\lda)\sin\left(\frac{\beta\phi}{2}\right)\sigma_1-ik_1(\lda)\cos\left(\frac{\beta\phi}{2}\right)\sigma_2\,,\\
\label{LP2}
V(x,t,\lda)&=&i\frac{\beta}{4}\Pi\,\sigma_3-ik_1(\lda)\sin\left(\frac{\beta\phi}{2}\right)\sigma_1-ik_0(\lda)\cos\left(\frac{\beta\phi}{2}\right)\sigma_2\,.
\eea
The three Pauli matrices are given as usual by 
\be
\sigma_1=\left(\begin{array}{cc}
0&1\\
1&0
\end{array}\right)~~,~~
\sigma_2=\left(\begin{array}{cc}
0&-i\\
i&0
\end{array}\right)~~,~~
\sigma_3=\left(\begin{array}{cc}
1&0\\
0&-1
\end{array}\right)\,.
\ee
The two function $k_0$ and $k_1$ read
\be
k_0(\lda)=\frac{m}{4}(\lda+\frac{1}{\lda})~~,~~k_1(\lda)=\frac{m}{4}(\lda-\frac{1}{\lda})\,.
\ee
The compatibility condition $\Psi_{xt}=\Psi_{tx}$ of the auxiliary problem is equivalent to the zero curvature condition
\be
U_t-V_x+[U,V]=0\,,
\ee
which in turn is equivalent to 
\be
\begin{cases}
\pi=\phi_t\,,\\
\Pi=-\phi_x\,,\\
\pi_t+\Pi_x+\frac{m^2}{\beta}\sin \beta\phi=0\,,
\end{cases}
\ee
and hence yields the sine-Gordon equation \eqref{sG}.

From the point of view of PDEs, an important characteristic of integrable equations is the existence of an infinite numbers of conservation laws and hence, of 
conserved quantities. Among many other things, the existence of a Lax pair formulation allows one to find these conservation laws easily and systematically. 
In Part Two, Chapter II.4 of \cite{FT}, such a systematic procedure is presented to extract the \textit{conserved quantities} (in time). 
Here, although we are strongly inspired by this derivation, we want to present a derivation of the \textit{conservation laws}
in a way that treats $x$ and $t$ on an equal footing, hence allowing us to extract conserved quantities in time and in space systematically. 
Another motivation is that conservation laws are more fundamental in the sense that conserved quantities are easily deduced from them by integration over 
an appropriate domain. 
Before we continue, we need to specify the chosen functional space and boundary conditions for the fields. When working in the traditional approach, 
with fixed $t$ and fields depending on $x$, we assume that 
\be
\label{space_BC}
\lim_{x\to\pm\infty}\phi(x)=\frac{2\pi Q_{\pm}}{\beta}=0\mod \frac{2\pi}{\beta}~~,~~\lim_{|x|\to\infty}\pi(x)=0\,,
\ee
where the boundary values are approached sufficiently fast \ie in the Schwarz sense. Similarly, when working in the new approach, 
with fixed $x$ and fields depending on $t$, we assume that 
\be
\label{time_BC}
\lim_{t\to\pm\infty}\phi(t)=\frac{2\pi \cQ_{\pm}}{\beta}=0\mod \frac{2\pi}{\beta}~~,~~\lim_{|t|\to\infty}\Pi(t)=0\,,
\ee
where the boundary values are approached sufficiently fast \ie in the Schwarz sense. Given our purposes, it is important to note that these two assumptions 
can coexist. Indeed, solving the initial value problem in the traditional approach, for initial data satisfying \eqref{space_BC}, 
naturally provides a large class of solutions, e.g. the $N$ soliton 
solutions, which are then defined for all $t\in\RR$ and satisfy \eqref{time_BC} at any fixed value of $x$.

Let us define 
\be
\Omega=e^{i\frac{\beta\phi}{4}\sigma_3}\,,
\ee
where we will treat $\Omega$ either as a function of $x$ for fixed time or vice versa, depending on the chosen picture.
Note that 
\bea
\lim_{x\to\pm\infty}U=(-1)^{Q_\pm}U_\infty & , & \lim_{t\to\pm\infty}U=(-1)^{\cQ_\pm}U_\infty\,,\\
\lim_{x\to\pm\infty}V=(-1)^{Q_\pm}V_\infty & , &\lim_{t\to\pm\infty}V=(-1)^{\cQ_\pm}V_\infty\,,
\eea
where,
\be
U_\infty=-ik_1(\lda)\sigma_2~~,~~V_\infty=-ik_0(\lda)\sigma_2\,.
\ee
So let us define
\be
N=\frac{1}{\sqrt{2}}(\1+i\sigma_1)\,,
\ee
and
\be
E_\pm(x,\lda)=e^{i\frac{\pi}{2}Q_\pm\sigma_3}\,N\,e^{-ik_1(\lda)x\sigma_3}~~,~~
\cE_\pm(t,\lda)=e^{i\frac{\pi}{2}\cQ_\pm\sigma_3}\,N\,e^{-ik_0(\lda)t\sigma_3}\,,
\ee
which satisfy
\be
\partial_x E_\pm(x,\lda)&=& (-1)^{Q_\pm}U_\infty\,E_\pm(x,\lda)\,,\\
\partial_t \cE_\pm(t,\lda)&=& (-1)^{\cQ_\pm}V_\infty\,\cE_\pm(t,\lda)\,.
\ee
Let $T(x,y,\lda)$ 
and ${\cal T}(t,\tau,\lda)$ denote respectively the space transition matrix (at fixed time) and the time transition matrix (at fixed position). 
They are solutions of $\Psi_x=U\Psi$ and $\Psi_t=V\Psi$ respectively and normalised by
\be
T(x,x,\lda)=\1~~,~~{\cal T}(t,t,\lda)=\1\,.
\ee
The corresponding monodromy matrices are 
\bea
T(\lda)&=&\lim_{\substack{x\to\infty\\y\to-\infty}}E_+^{-1}(x,\lda)T(x,y,\lda)E_-(y,\lda)\,,\\
\cT(\lda)&=&\lim_{\substack{t\to\infty\\\tau\to-\infty}}\cE_+^{-1}(t,\lda)\cT(t,\tau,\lda)\cE_-(\tau,\lda)\,.
\eea
For the purpose of discussing conservations laws and conserved quantities, it is convenient to gauge away the charges at $\pm\infty$ and 
to consider the following gauge transformed matrices
\be
T(x,y,\lda)=\Omega(x)\widehat{T}(x,y,\lda)\Omega(y)^{-1}~~,~~{\cal T}(t,\tau,\lda)=\Omega(t)\widehat{{\cal T}}(t,\tau,\lda)\Omega(\tau)^{-1}\,.
\ee
Then $\widehat{T}$ is a solution of $\Psi_x=\widehat{U}\,\Psi$ where 
\be
\widehat{U}=-i\frac{\beta}{4}(\phi_x+\pi)\sigma_3-i\lda\frac{m}{4}\, \sigma_2\,+i\frac{m}{4\lda}\Omega^{-2}\, \sigma_2\,\Omega^2\,,
\ee
and $\widehat{{\cal T}}$ is a solution of $\Psi_t=\widehat{V}\,\Psi$ where 
\be
\widehat{V}=-i\frac{\beta}{4}(\phi_t-\Pi)\sigma_3-i\lda\frac{m}{4}\, \sigma_2\,-i\frac{m}{4\lda}\Omega^{-2}\, \sigma_2\,\Omega^2\,.
\ee
Observe that 
\bea
\lim_{x\to\pm\infty}\widehat{U}=U_\infty=\lim_{t\to\pm\infty}\widehat{U}~~,~~
\lim_{x\to\pm\infty}\widehat{V}=V_\infty=\lim_{t\to\pm\infty}\widehat{V}\,.
\eea
Accordingly, we introduce 
\be
E_0(x,\lda)=N\,e^{-ik_1(\lda)x\sigma_3}~~,~~
\cE_0(t,\lda)=N\,e^{-ik_0(\lda)t\sigma_3}\,,
\ee
and the following half-infinite transition matrices
\be
\label{half_inf}
\widehat{T}_\pm(x,t,\lda)=\lim_{y\to\pm\infty}\widehat{T}(x,y,t,\lda)E_0(y,\lda)~~,~~\widehat{{\cal T}}_\pm(x,t,\lda)=\lim_{\tau\to\pm\infty}
\widehat{{\cal T}}(t,\tau,x,\lda)\cE_0(\tau,\lda)\,.
\ee
Finally, note that the corresponding monodromy matrices
\bea
\label{def_mono_space}
\widehat{T}(\lda)&=&\lim_{\substack{x\to\infty\\y\to-\infty}}E_0^{-1}(x,\lda)\widehat{T}(x,y,\lda)E_0(y,\lda)\,,\\
\label{def_mono_time}
\widehat{\cT}(\lda)&=&\lim_{\substack{t\to\infty\\\tau\to-\infty}}\cE_0^{-1}(t,\lda)\widehat{\cT}(t,\tau,\lda)\cE_0(\tau,\lda)\,,
\eea
are equal to $T(\lda)$ and $\cT(\lda)$ respectively, so that they capture the same information about the system.
Using the the zero curvature condition, we can study the time evolution of the transition matrix $\widehat{T}(x,y,\lda)$ 
and the space evolution of the transition matrix $\widehat{{\cal T}}(t,\tau,\lda)$. Writing the time and space dependence explicitely, we get
\bea
\label{time_evo}
\partial_t\,\widehat{T}(x,y,t,\lda)&=&\widehat{V}(x,t,\lda)\widehat{T}(x,y,t,\lda)-\widehat{T}(x,y,t,\lda)\widehat{V}(y,t,\lda)\\
\label{space_evo}
\partial_x\,\widehat{{\cal T}}(t,\tau,x,\lda)&=&\widehat{U}(x,t,\lda)\widehat{{\cal T}}(t,\tau,x,\lda)-\widehat{{\cal T}}(t,\tau,x,\lda)\widehat{U}(x,\tau,\lda)\,.
\eea
These equations allows us to deduce the time evolution of $\widehat{T}(\lda)$ and the space evolution of $\widehat{\cT}(\lda)$,
\bea
\label{time_evo_mono}
\partial_t \widehat{T}(t,\lda)&=&-ik_0(\lda)[\sigma_3,\widehat{T}(t,\lda)]\,,\\
\label{space_evo_mono}
\partial_x \widehat{\cT}(x,\lda)&=&-ik_1(\lda)[\sigma_3,\widehat{\cT}(x,\lda)]\,.
\eea
This shows that the $(1,1)$ entry of $\widehat{T}(\lda)$ is time-independent and we denote it $a(\lda)$. Similarly, the
the $(1,1)$ entry of $\widehat{\cT}(\lda)$ is space-independent and we denote it $\fa(\lda)$. 

The key to deriving the conservation laws is the following result that we prove in Appendix \ref{AppA}
\be
\label{equality}
\widehat{T}_-(x,t,\lda)e^{-ik_0t\sigma_3}=\widehat{\cT}_-(x,t,\lda)e^{-ik_1x\sigma_3}\,.
\ee
We denote the common value by 
\be
R(x,t,\lda)e^{-i(k_1x+k_0t)\sigma_3}\,,
\ee
which is then the solution of 
\be
\begin{cases}
\partial_xR=\widehat{U}R+ik_1R\sigma_3\,,\\
\partial_tR=\widehat{V}R+ik_0R\sigma_3\,,
\end{cases}
\ee
satisfying
\be
\lim_{x\to-\infty}R(x,t,\lda)=N=\lim_{t\to-\infty}R(x,t,\lda)\,.
\ee
Generalising the standard approach to the present setting, we can extract the conservation laws as $\lda\to\infty$ by writing 
\be
R(x,t,\lda)=\frac{1}{\sqrt{2}}(\1+\Gamma(x,t,\lda))e^{Y(x,t,\lda)}\,,
\ee
where $Y$ is a diagonal matrix with expansion 
\be
\label{expansion_Y}
Y(x,t,\lda)=i\sum_{n=1}^\infty\frac{Y_n(x,t)}{\lda^n}\,,
\ee
and $\Gamma$ is an off-diagonal matrix with expansion
\be
\label{expansion_Gamma}
\Gamma(x,t,\lda)=\sum_{n=0}^\infty\frac{\Gamma_n(x,t)}{\lda^n}\,.
\ee
As a consequence, dropping the arguments of the functions for conciseness, we derive the following equations
\bea
\label{gen_x}
Y_x&=&\widehat{U}_d+\widehat{U}_o\Gamma+ik_1(\lda)\sigma_3\,,\\
\label{gen_t}
Y_t&=&\widehat{V}_d+\widehat{V}_o\Gamma+ik_0(\lda)\sigma_3\,,
\eea
as well as the associated space and time Riccati equations for $\Gamma$
\bea
\label{Riccati_x}
\Gamma_x&=&\widehat{U}_o+\widehat{U}_d\Gamma-\Gamma\widehat{U}_d-\Gamma\widehat{U}_o\Gamma\,,\\
\label{Riccati_t}
\Gamma_t&=&\widehat{V}_o+\widehat{V}_d\Gamma-\Gamma\widehat{V}_d-\Gamma\widehat{V}_o\Gamma\,,
\ee
where the subscripts $d$ and $o$ denote the diagonal and off-diagonal parts of the matrices respectively.
At this stage, the conservation laws are obtained by cross differentiation of \eqref{gen_x} and \eqref{gen_t}
\be
\label{conservation_law}
\left(\widehat{U}_d+\widehat{U}_o\Gamma+ik_1(\lda)\sigma_3\right)_t=
\left(\widehat{V}_d+\widehat{V}_o\Gamma+ik_0(\lda)\sigma_3\right)_x\,.
\ee
In particular, integrating in $x$ from $-\infty$ (where $Y=0$) to $\infty$, 
\eqref{conservation_law} yields
\be
\partial_t\left(\lim_{x\to\infty}Y(x,t,\lda)\right)=\lim_{\substack{x\to\infty\\y\to-\infty}}\left[\widehat{V}_d+\widehat{V}_o\Gamma+ik_0(\lda)\sigma_3\right]_y^x
=0\,.
\ee
We now show that this is of course consistent with \eqref{time_evo_mono} which shows that the diagonal part of $\widehat{T}(\lda)$ is time independent.
In the process, we also recover the well-known fact that 
$\ln a(\lda)$ generates conserved quantities and, as $\lda\to\infty$,
\be
\label{expansion_lna}
\ln a(\lda)=i\sum_{n=1}^\infty\frac{I_n}{\lda^n}\,,
\ee
where here,
\be
I_n=\lim_{x\to\infty} y_n(x,t)\,,
\ee
where $y_n$ is the $(1,1)$ entry of $Y_n$ in \eqref{expansion_Y}.
Inserting \eqref{expansion_Gamma} into \eqref{Riccati_x}, we find $\Gamma_0=i\sigma_1$ and
\bea
\Gamma_{n+1}&=&-\frac{2i}{m}\Gamma_{nx}\sigma_3-\frac{\beta}{m}(\pi+\phi_x)\Gamma_n-\frac{i}{2}\sigma_1(e^{i\beta\phi\sigma_3}-e^{-i\beta\phi\sigma_3})\delta_{n,1}
\\
&&+\frac{i}{2}\left(\sum_{p=1}^n\Gamma_p\,\sigma_1\,\Gamma_{n+1-p}-\sum_{p=0}^{n-1}\Gamma_p\,\sigma_1\,e^{i\beta\phi\sigma_3}\,\Gamma_{n-1-p}\right)\,.
\eea
Then, \eqref{gen_x} yields
\be
Y_n(x,t)=-\frac{m}{4}\int_{-\infty}^x\sigma_2\left(\Gamma_{n+1}(\xi,t)-e^{i\beta\phi(\xi,t)\sigma_3}\Gamma_{n-1}(\xi,t)+i\sigma_1\,\delta_{n,1}\right)d\xi\,.
\ee
Hence, noting that for $n\ge 1$, $\Gamma_n(x,t)\to 0$ as $x\to-\infty$, we find, as $\lda\to\infty$,
\be
\widehat{T}(\lda)=e^{P(\lda)}\,,
\ee
where 
\be
P(\lda)=\lim_{x\to\infty}Y(x,t,\lda)\,,
\ee
as required. In particular, as $\lda\to\infty$, $\ln a(\lda)$ coincides with the $(1,1)$ entry of 
$\displaystyle \lim_{x\to\infty}Y(x,t,\lda)$, which is indeed time independent, and \eqref{expansion_lna} holds.
To complete the analysis, one has to study the behaviour of the previous quantities as $\lda\to 0$. This is largely simplified by noting that if $\Gamma(x,t,\lda)$ 
is solution of \eqref{Riccati_x} then $e^{i\beta\phi\sigma_3}\Gamma(x,t,-\frac{1}{\lda})$ is solution of \eqref{Riccati_x} with $\phi$ changed 
to $-\phi$ everywhere. Therefore, we deduce that, as $\lda\to 0$,
\be
R(x,t,\lda)=\frac{1}{\sqrt{2}}(\1+\Gamma(x,t,\lda))e^{Y(x,t,\lda)}\,,
\ee
with
\be
\Gamma(x,t,\lda)=\sum_{n=0}^\infty (-1)^n\Gamma^{'}_n(x,t)\lda^n\,,
\ee
and $\Gamma^{'}_n$ is obtained from $\Gamma_n$ by changing $\phi$ to $-\phi$. Finally, we obtain
\be
Y(x,t,\lda)=i\sum_{n=0}^\infty Y^{'}_n(x,t)\lda^n\,,
\ee
with
\be
Y^{'}_0(x,t)=-\frac{\beta}{2}\int_{-\infty}^x\partial_\xi \phi(\xi,t)\,d\xi\,,
\ee
and, for $n\ge 1$,
\be
Y^{'}_n(x,t)=-\frac{m}{4}\int_{-\infty}^x \sigma_2\left((-1)^{n-1}e^{-\beta\phi(\xi,t)\sigma_3}\Gamma^{'}_{n-1}(\xi,t)-
(-1)^{n+1}\Gamma^{'}_{n+1}(\xi,t)-i\sigma_1\,\delta_{n,1}\right)\,d\xi\,,
\ee
Therefore, as $\lda\to 0$, one can deduce that 
\be
\ln a(\lda)=i\sum_{n=0}^\infty I_{-n}\lda^n
\ee
where
\be
I_{-n}=\lim_{x\to\infty}y^{'}_n(x,t)\,,
\ee 
and $y^{'}_n$ is the $(1,1)$ entry of $Y^{'}_n$.
In particular, an explicit calculation yields
\be
I_{-1}-I_1=\frac{\beta^2}{2m}\int_{-\infty}^\infty\left[\frac{1}{2}(\pi^2+\phi_x^2)+\frac{m^2}{\beta^2}(1-\cos\beta\phi)\right]dx=\frac{\beta^2}{2m}H_S\,,
\ee
where $H_S$ is the Hamiltonian in \eqref{Hams}.

Similarly, integrating \eqref{conservation_law} in $t$ from $-\infty$ (where $Y=0$) to $\infty$, we can perform an analogous analysis 
in the new picture and extract local conserved quantities in space. 
Indeed, 
\be
\partial_x\left(\lim_{t\to\infty}Y(x,t,\lda)\right)=\lim_{\substack{t\to\infty\\\tau\to-\infty}}\left[\widehat{U}_d+\widehat{U}_o\Gamma+ik_1(\lda)\sigma_3\right]_\tau^t
=0\,.
\ee
In this second picture, one should use \eqref{Riccati_t} and \eqref{gen_t} instead of \eqref{Riccati_x} and \eqref{gen_x} 
to study the asymptotic behaviour as $\lda\to\infty$ and $\lda\to 0$ and extract the conserved quantities. Doing so, as
$\lda\to\infty$ we find 
\be
\Gamma(x,t,\lda)=\sum_{n=0}^\infty\frac{\gamma_n(x,t)}{\lda^n}\,,
\ee
with $\gamma_0=i\sigma_1$ and
\bea
\label{time_recursion}
\gamma_{n+1}&=&-\frac{2i}{m}\gamma_{nt}\sigma_3-\frac{\beta}{m}(\phi_t-\Pi)\gamma_n+\frac{i}{2}\sigma_1(e^{i\beta\phi\sigma_3}-e^{-i\beta\phi\sigma_3})\delta_{n,1}
\nonumber\\
&&+\frac{i}{2}\left(\sum_{p=1}^n\gamma_p\,\sigma_1\,\gamma_{n+1-p}+\sum_{p=0}^{n-1}\gamma_p\,
\sigma_1\,e^{i\beta\phi\sigma_3}\,\gamma_{n-1-p}\right)\,,
\eea
as well as 
\be
Y(x,t,\lda)=i\sum_{n=1}^\infty\frac{\cY_n(x,t)}{\lda^n}\,,
\ee
where
\be
\cY_n(x,t)=-\frac{m}{4}\int_{-\infty}^t\sigma_2\left(\gamma_{n+1}(x,\tau)+e^{i\beta\phi(x,\tau)\sigma_3}\gamma_{n-1}(x,\tau)-i\sigma_1\,\delta_{n,1}\right)d\tau.
\ee
Therefore, in view of \eqref{def_mono_time} and the fact that $\gamma_n(x,t)\to 0$ as $t\to-\infty$ for $n\ge 1$, we obtain that as $\lda\to\infty$, 
\be
\ln\fa(\lda)=i\sum_{n=1}^\infty\frac{J_n}{\lda^n}\,,
\ee
where
\be
J_n=\lim_{t\to\infty}\Upsilon_n(x,t)\,,
\ee
and $\Upsilon_n$ is the $(1,1)$ entry of $\cY_n$.
So here again, in the new picture, we obtain that $\ln \fa(\lda)$ generates conserved quantities (in space) $J_n$.
To obtain the behaviour as $\lda\to 0$, we make a similar observation as before to relate it to the behaviour as $\lda\to\infty$. In the present 
case, we find that as $\lda\to 0$
\be
\Gamma(x,t,\lda)=e^{-i\beta\phi(x,t)\sigma_3}\Gamma^{'}(x,t,\frac{1}{\lda})\,,
\ee
where the dot means that $\Gamma^{'}$ solves \eqref{Riccati_t} with $\phi$ changed to $-\phi$. Hence, we 
deduce
\be
\Gamma(x,t,\lda)=e^{-i\beta\phi(x,t)\sigma_3}\sum_{n=0}^\infty \gamma^{'}_n(x,t)\lda^n\,,
\ee
where $\gamma^{'}_n$ is $\gamma_n$ with $\phi\to-\phi$. Accordingly, as $\lda\to 0$,
\be
Y(x,t,\lda)=i\sum_{n=0}^\infty \cY_n^{'}(x,t)\lda^n\,,
\ee
with
\bea
\cY_0^{'}(x,t)&=&-\frac{\beta}{2}\int_{-\infty}^t\phi_\tau(x,\tau)\,d\tau\,,\\
\cY_n^{'}(x,t)&=&-\frac{m}{4}\int_{-\infty}^t\sigma_2\left[e^{-i\beta\phi(x,\tau)\sigma_3}\gamma^{'}_{n-1}(x,\tau)+\gamma^{'}_{n+1}(x,\tau)-i\sigma_1\,\delta_{n,1} \right]d\tau
\,,n\ge 1\,.
\eea
In particular, as $\lda\to 0$,
\be
\ln \fa(\lda)=i\sum_{n=0}^\infty J_{-n}\lda^n
\ee
where
\be
J_{-n}=\lim_{t\to\infty}\Upsilon^{'}_n(x,t)
\ee
and $\Upsilon^{'}_n$ is the $(1,1)$ entry of $\cY^{'}_n$. As an important check, an explicit calculation yields the new Hamiltonian $H_T$ from 
\eqref{Ham_T_density},\eqref{Hams}, as the 
following combination of integrals
\be
J_{1}+J_{-1}=-\frac{\beta^2}{2m}\int_{-\infty}^\infty\left[\frac{1}{2}(\Pi^2+\phi_t^2)-\frac{m^2}{\beta^2}(1-\cos\beta\phi)\right]dx=\frac{\beta^2}{2m}H_T\,.
\ee

Let us summarize the results for clarity. Writing,
\be
I(\lda)=\int_{-\infty}^\infty(\widehat{U}_d+\widehat{U}_o\Gamma+ik_1(\lda)\sigma_3)dx\,,
\ee
then
\be
\partial_t\,I(\lda)=0\,.
\ee
Using the space Riccati equation \eqref{Riccati_x} for $\Gamma$, one extracts conserved quantities in time by inserting an expansion of $\Gamma$ as $\lda\to\infty$ 
and $\lda\to 0$. This gives a set $I_n$, $n\in\ZZ$ and usual charges are linear combinations of these quantities, e.g. the Hamiltonian $H_S$ is proportional 
to $I_{-1}-I_1$ and the topological charge $Q_+-Q_-$ is proportional to $I_0$. The $(1,1)$ entry $a(\lda)$ of the monodromy matrix $\widehat{T}(\lda)=T(\lda)$ 
generates the $I_n$'s according to
\bea
\ln a(\lda)&=&i\sum_{n=1}^\infty\frac{I_n}{\lda^n}~~,~~\lda\to\infty\,,\\
\ln a(\lda)&=&i\sum_{n=0}^\infty I_{-n}\lda^n~~,~~\lda\to 0\,.
\eea
In the new picture, writing,
\be
J(\lda)=\int_{-\infty}^\infty(\widehat{V}_d+\widehat{V}_o\Gamma+ik_0(\lda)\sigma_3)dx\,,
\ee
then
\be
\partial_x\,J(\lda)=0\,.
\ee
Using the time Riccati equation \eqref{Riccati_t} for $\Gamma$, one extracts conserved quantities in space by inserting an expansion of $\Gamma$ as $\lda\to\infty$ 
and $\lda\to 0$. This gives a set $J_n$, $n\in\ZZ$ and relevant charges in the new picture are linear combinations of these quantities, e.g. the Hamiltonian $H_T$ is proportional 
to $J_{-1}+J_1$ and the topological charge $\cQ_+-\cQ_-$ is proportional to $J_0$. The $(1,1)$ entry $\fa(\lda)$ of the monodromy matrix $\widehat{\cT}(\lda)=\cT(\lda)$ 
generates the $J_n$'s according to
\bea
\ln \fa(\lda)&=&i\sum_{n=1}^\infty\frac{J_n}{\lda^n}~~,~~\lda\to\infty\,,\\
\ln \fa(\lda)&=&i\sum_{n=0}^\infty J_{-n}\lda^n~~,~~\lda\to 0\,.
\eea

\subsection{Classical $r$-matrix approach for the two Poisson brackets}\label{two_app}

The combination of the Lax pair approach to conserved quantities together with the Hamiltonian approach to integrable field theories culminates in the 
so-called classical $r$-matrix approach \cite{Sklyanin}. In our context, it gives a convenient way of combining the 
multisymplectic Hamiltonian approach to the model with the existence of an infinite hierarchy of conserved quantities. 
The latter appear as functions that are in involution with respect to the Poisson bracket used to describe the model. The novelty here is 
the classical $r$ matrix approach in terms of the new bracket $\{~,~\}_T$ and the associated role of the conserved quantities in space encoded 
in $\fa(\lda)$, the $(1,1)$ entry of the time monodromy matrix $\cT(\lda)$.

\subsubsection{The standard approach with $\{~,~\}_S$}\label{standard}

Recall that $U$ in \eqref{LP1} is given by
\be
U(x,\lda)&=&-i\frac{\beta}{4}\pi(x)\sigma_3-ik_0(\lda)\sin\left(\frac{\beta\phi(x)}{2}\right)\sigma_1-ik_1(\lda)\cos\left(\frac{\beta\phi(x)}{2}\right)\sigma_2\,.
\ee
where we have dropped the time variable since it assumed to be fixed at some initial value in this approach.
Then, following a standard calculation (see e.g. \cite{FT}), one finds
 \bea 
 \label{ultraloc}
 \{U_1(x,\lda),U_2(y,\mu)\}_S=\delta(x-y)\left[r(\lda,\mu),U_1(x,\lda)+ U_2(y,\mu)\right]\,, 
 \eea
 where we have used the notation $U_1=U\otimes \1$, $U_2=\1\otimes U$ and $r(\lda,\mu)$ 
 is the classical $r$-matrix given by
 \bea
 \label{r_matrix}
 r(\lda,\mu)=f(\lda,\mu)(\1_2\otimes \1_2-\sigma_3\otimes \sigma_3)+g(\lda,\mu)(\sigma_1\otimes \sigma_1+\sigma_2\otimes \sigma_2)\,,
 \eea
with
\be
f(\lda,\mu)=-\gamma\frac{\lda^2+\mu^2}{\lda^2-\mu^2}~~,~~g(\lda,\mu)=2\gamma\frac{\lda\mu}{\lda^2-\mu^2}~~,~~\gamma=\frac{\beta^2}{16}\,.
\ee 
Setting 
$\lda=e^{i\alpha}$, $\mu=e^{i\beta}$, it is conveniently rewritten in trigonometric form where it only depends on the difference $\alpha-\beta$ and
\be
r(\alpha)=\frac{i\gamma}{\sin\alpha}\left(\begin{array}{cccc}
0&0&0&0\\
0&\cos\alpha&-1&0\\
0&-1&\cos\alpha&0\\
0&0&0&0
\end{array}\right)\,.
\ee
For the transition matrix $T(x,y,\lda)$, $y<x$, one then finds
 \be
  \{T_1(x,y,\lda),T_2(x,y,\mu)\}_S=\left[r(\lda,\mu),T(x,y,\lda)\otimes T(x,y,\lda)\right]\,, 
 \ee
 Using definition \eqref{def_mono_space} for the monodromy matrix and a careful limiting procedure (see Section II.6 of \cite{FT}), one obtains the 
 infinite volume Poisson brackets for the monodromy matrix $T(\lda)$ as
 \be
 \{T_1(\lda),T_2(\mu)\}_S=r_+(\lda,\mu)\,T_1(\lda)\,T_2(\mu)-T_1(\lda)\,T_2(\mu)\,r_-(\lda,\mu)\,,
 \ee 
 where 
 \be
 \label{infinite_vol_r}
 r_\pm(\lda,\mu)=-\frac{\gamma}{2}
 \left(
\begin{array}{cccc}
\frac{\lda-\mu}{\lda+\mu} &0 & 0& 0\\
0& p.v.\frac{\lda+\mu}{\lda-\mu} & \mp i\pi(\lda+\mu)\delta(\lda-\mu) & 0\\
0 & \pm i\pi(\lda+\mu)\delta(\lda-\mu) & p.v.\frac{\lda+\mu}{\lda-\mu}  & 0\\
0 & 0& 0 & \frac{\lda-\mu}{\lda+\mu}
\end{array}\right)\,.
 \ee
One can then conclude that $a(\lda)$ Poisson commutes with $a(\mu)$
\be
\{a(\lda),a(\mu)\}_S=0\,.
\ee 
This is the key relation showing that the integrals of motion $I_n$, n$\in\ZZ$ are in involution with respect to $\{~,~\}_S$. 
This is taken as the definition of Liouville integrability of the sine-Gordon model.

\subsubsection{Classical $r$-matrix approach for the new bracket $\{~,~\}_T$}\label{class_new}

In view of the results obtained in Section \ref{multi_sG}, one can easily derive a treatment of the classical $r$-matrix approach to sine-Gordon with 
respect to new Poisson bracket $\{~,~\}_T$ that is completely analogous to the standard one reviewed previously.
Starting with
\be
\label{LPUV}
V(t,\lda)=i\frac{\beta}{4}\Pi(t)\sigma_3-ik_1(\lda)\sin\left(\frac{\beta\phi(t)}{2}\right)\sigma_1-ik_0(\lda)\cos\left(\frac{\beta\phi(t)}{2}\right)\sigma_2
\ee  
and using \eqref{time_brackets}, one obtains the following result by direct calculation.
\begin{proposition} Let the
Poisson bracket $\{~,~\}_T$ be given by \eqref{time_brackets} and $V$ be
given by \eqref{LPUV}.  Then, 
\bea
\{V_1(t,\lda),V_2(\tau,\mu)\}_T=-\delta(t-\tau)
\left[r(\lda,\mu),V_1(t,\lda)+ V_2(\tau,\mu)\right] \,, \eea with the
same classical $r$-matrix as in \eqref{r_matrix}.  
\end{proposition} 
As a direct consequence, we obtain for the transition matrix the following 
\begin{corollary}
\label{Coro_Mt}
 For $\tau<t$, 
\bea 
\label{PBT}
\{\cT_1(t,\tau,\lda),\cT_2(t,\tau,\mu)\}_T=-\left[r(\lda,\mu),\cT(t,\tau,\lda)\otimes \cT(t,\tau,\mu)\right]\,,
\eea 
and, 
 \be
 \{\cT_1(\lda),\cT_2(\mu)\}_T=-r_+(\lda,\mu)\,\cT_1(\lda)\,\cT_2(\mu)+\cT_1(\lda)\,\cT_2(\mu)\,r_-(\lda,\mu)\,,
 \ee 
 where $r_\pm$ is given by \eqref{infinite_vol_r}. In particular,
\be
\{\fa(\lda),\fa(\mu)\}_T=0\,.
\ee
\end{corollary} 
We now get that the integrals $J_n$, $n\in\ZZ$ generated by $\ln \fa(\lda)$ are in involution with respect to $\{~,~\}_T$. 
In this picture, this fact can be taken as a definition of Liouville integrability of the model which is alternative (and equivalent) to the 
standard one given in the previous section. From the point of view of the classical $r$ matrix, the two points of view (space or time) 
are once again equivalent.

\section{Sine-Gordon model with a defect: Liouville integrability}
\label{defField}

\subsection{Defect conditions as "frozen" B\"acklund transformations}

\subsubsection{Review of the Lagrangian approach}

 Viewing a defect in space as an internal boundary condition on the fields and
their time and space derivatives at a given point, the fruitful idea of
{\it frozen} B\"acklund transformations,
 originally noticed in \cite{BCZ},
 is a convenient way of introducing integrable defects in classical field theories described by a Lax pair.  
 Initially, starting from a Lagrangian density of the form
 \be
 \label{Lagrangian}
 \cL=\theta(-x)\cL_{\widetilde{\phi}}+\theta(x)\cL_\phi-\delta(x)\left(\frac{1}{2}(\widetilde{\phi}\phi_t-\phi\widetilde{\phi}_t)-\cB\right)\,,
 \ee
 where, without loss of generality, the location of the defect has been chosen to be $x=0$, it was required that the defect conditions 
 between the sine-Gordon fields $\phi$ and $\widetilde{\phi}$ on either side of the defect were such that the associated modified momentum 
 was conserved in time. This led to a solution for $\cB$ of the form
 \be
 \label{form_B}
 \cB=\frac{2m}{\beta^2}\left(\sigma\cos\beta\left(\frac{\widetilde{\phi}+\phi}{2}\right)+\sigma^{-1}\cos\beta\left(\frac{\widetilde{\phi}-\phi}{2}\right)\right)
 \ee
 and to the defect conditions, at $x=0$,
 \be
 \label{defect_conditions}
\begin{cases}
 \widetilde{\phi}_x-\phi_t=\frac{m}{\beta}\left(\sigma\sin\beta\left(\frac{\widetilde{\phi}+\phi}{2}\right)+\sigma^{-1}\sin\beta\left(\frac{\widetilde{\phi}-\phi}{2}\right)\right)\,,\\
 \widetilde{\phi}_t-\phi_x=\frac{m}{\beta}\left(\sigma\sin\beta\left(\frac{\widetilde{\phi}+\phi}{2}\right)-\sigma^{-1}\sin\beta\left(\frac{\widetilde{\phi}-\phi}{2}\right)\right)\,.
 \end{cases}
 \ee
These conditions also ensured that the modified Hamiltonian and first higher integral were also conserved. They were recognized as frozen B\"acklund transformations 
of the sine-Gordon model at the location of the defect. 

\subsubsection{Review of the (traditional) Lax pair approach}

Later, this observation was exploited in \cite{VC} for a systematic derivation of the generating function 
of the modified conserved quantities for all models in the AKNS hierarchies \cite{AKNS}. In particular, the sine-Gordon model in light cone coordinates
was discussed there. Here, for our purposes, we present the procedure of \cite{VC} but for sine-Gordon in laboratory coordinates. The central ingredient is 
the so-called B\"acklund or defect matrix $L$ which is required to be a solution of 
\be
\label{eq_L}
L_t=VL-L\widetilde{V}~~,~~x=0\,,
\ee
where $V$ is the time Lax matrix \eqref{LP2} and $\widetilde{V}$ the same time Lax matrix with $\phi,\Pi$ replaced by $\widetilde{\phi},\tPi$. With an appropriate solution 
of $L$, on the one hand one reproduces the defect conditions \eqref{defect_conditions} and, on the other hand, 
$L$ can be used to connect the transition matrices of both theories and obtain the monodromy matrix of the theory on the full line with 
a defect at $x=0$. Assuming that the fields $\phi,\pi,\Pi$ describe the model for $x>0$ and the fields $\tphi,\tpi,\tPi$ 
describe the theory for $x<0$, in the notations of Section \ref{CL}, we see that $\widehat{T}_+^{-1}(0,t,\lda)$ describes the positive half-line $(0,\infty)$ 
at time $t$ while $\widehat{\widetilde{T}}_-(0,t,\lda)$ describes the negative half-line $(-\infty,0)$ at time $t$. Therefore, the system on the line with a 
defect encoded in $L$ is described by the following monodromy matrix
\be
\label{def_M_S}
\cM_S(t,\lda)=\widehat{T}_+^{-1}(0,t,\lda)\,\widehat{L}(t,\lda)\,\widehat{\widetilde{T}}_-(0,t,\lda)\,,
\ee
where 
\be
\widehat{L}(t,\lda)=\Omega(t)^{-1} L(t,\lda) \widetilde{\Omega}(t)
\ee
is the gauged B\"acklund/defect matrix. A direct computation then yields
\be
\partial_t \cM_S(t,\lda)=-ik_0[\sigma_3,\cM_S]\,,
\ee
which show that the diagonal part of $\cM_S$ is automatically time independent for any solution $L$ of \eqref{eq_L}. Let us 
denote by
\be
\cI(\lda)=\ln(\cM_{S})_d\,,
\ee
the logarithm of the diagonal part of $\cM_S$. We recover the analog of the results of \cite{VC} in the following form
\be
\cI(\lda)=I_+(\lda)+\widetilde{I}_-(\lda)+I_{defect}(\lda)
\ee
where
\be
I_+(\lda)=\int_0^\infty\left(\widehat{U}_d-\Gamma\widehat{U}_o+ik_1\sigma_3\right)dx~~,~~
\widetilde{I}_-(\lda)=\int_{-\infty}^0\left(\widehat{\widetilde{U}}_d-\widetilde{\Gamma}\widehat{\widetilde{U}}_o+ik_1\sigma_3\right)dx\,,
\ee
and
\be
I_{defect}(\lda)=\ln\frac{1}{2}\left(\widehat{L}_d-\Gamma\widehat{L}_o+\widehat{L}_o\widetilde{\Gamma}-\Gamma\widehat{L}_d\widetilde{\Gamma}\right)\,.
\ee
As explained in Section \ref{CL}, in this picture, $\Gamma$, resp. $\widetilde{\Gamma}$, is determined from the space Riccati equation \eqref{Riccati_x} involving 
$\widehat{U}$, resp. $\widehat{\widetilde{U}}$. 

These are general results valid for any B\"acklund matrix satisfying \eqref{eq_L}. In the present case, the connection with the defect conditions \eqref{defect_conditions} goes as follows. We 
found that the following solution for $L$ is such that \eqref{eq_L} is equivalent to \eqref{defect_conditions},
\be
\label{form_L}
L(t,\lda)=\Omega\widetilde{\Omega}^{-1}-\frac{i\sigma}{\lda}\widetilde{\Omega}^{-1}\sigma_2\Omega\,,
\ee
where all the fields $\phi$ and $\tphi$ are understood as depending on $t$ only for fixed $x=0$.
The quantity $I_{defect}(\lda)$
is the generating matrix of the so-called defect contributions to the hierarchy of conserved quantities for the model with an integrable defect given by the 
condition \eqref{eq_L}. In the special case where $L$ is given by \eqref{form_L}, it gives the defect contributions corresponding to the defect conditions 
\eqref{defect_conditions}. Of course, it can be checked that the present results agree with the results obtained in \cite{BCZ,VC}.

\subsubsection{Lax pair approach in the new picture}\label{Lax_new}

In the new picture, for $x>0$, the system is described by the time monodromy matrix $\widehat{\cT}(x,\lda)$ defined as in \eqref{def_mono_time} For $x<0$,
it is described by the time monodromy matrix $\widehat{\widetilde{\cT}}(x,\lda)$, defined similarly, but with $\phi,\pi,\Pi$ replaced by $\tphi,\tpi,\tPi$. 
Recall that the space evolution of these monodromy matrices is given by \eqref{space_evo_mono}.
The 
theory on the positive half-line is assumed to be continuous as $x\to 0^+$ and the theory on the negative half-line is assumed to 
be continuous as $x\to 0^-$. At $x=0$, the defect conditions \eqref{defect_conditions} relating the fields on each side, and 
encoded in \eqref{eq_L}, translate into the following connection formula for the time transition matrices $\widehat{\cT}(t,\tau,0,\lda)$ and 
$\widehat{\widetilde{\cT}}(t,\tau,0,\lda)$
\be
\widehat{\cT}(t,\tau,0,\lda)=\widehat{L}(t,\lda)\widehat{\widetilde{\cT}}(t,\tau,0,\lda)\widehat{L}(\tau,\lda)^{-1}\,.
\ee
In turn, this gives,
\be
\widehat{\cT}(0,\lda)=B_+(\lda)\widehat{\widetilde{\cT}}(0,\lda)B_-^{-1}(\lda)\,,
\ee
where
\be
\label{L_infinity}
B_\pm(\lda)=\frac{\lda}{\lda+i\sigma}\left(\1-\frac{i\sigma}{\lda}(-1)^{\widetilde{\cQ}_\pm+\cQ_\pm}\sigma_3\right)\,.
\ee
In particular, the generating functions of the conserved quantities are related by
\be
\label{transfo_gen_integrals}
\ln\fa(\lda)=\ln\widetilde{\fa}(\lda)+\ln C(\lda)\,,
\ee
where
\be
C(\lda)=\frac{1}{\lda+i\sigma}(\lda-i\sigma(-1)^{\widetilde{\cQ}_++\cQ_+})(\lda+i\sigma(-1)^{\widetilde{\cQ}_-+\cQ_-})\,.
\ee 
Hence, we find that the new Hamiltonians $H_T$ and $\widetilde{H}_T$ are related by 
\be
H_T=\widetilde{H}_T+\frac{2m}{\beta^2}(\sigma+\frac{1}{\sigma})\left((-1)^{\widetilde{\cQ}_++\cQ_+}-(-1)^{\widetilde{\cQ}_-+\cQ_-}\right)\,.
\ee

\subsection{Defect conditions as canonical transformations}\label{defect_cano}

We are now ready to tackle the crux of the matter and to show that the integrable defect conditions for sine-Gordon considered so far in the literature, and 
given here by \eqref{defect_conditions}, are in fact canonical transformations between the fields $(\phi,\Pi)$ and $(\tphi,\tPi)$ with respect to 
the new Poisson structure given by $\{~,~\}_T$.

The idea is to adapt the results obtained in \cite{KW,K} to the present multisymplectic approach. In doing so, we obtain a new interpretation of the
defect density in the Lagrangian \eqref{Lagrangian} 
\be
\cL_{defect}=\frac{1}{2}(\widetilde{\phi}\phi_t-\phi\widetilde{\phi}_t)-\cB
\ee
as the density for the generating functional of the canonical transformation correponding to the defect conditions. Let us first review 
briefly the idea of \cite{KW,K}. Given that one deals with an integrable model, one consider transformations that leave not only the form of Hamilton's equations invariant but also the form 
of all the integrals of motion. The usual requirement is that the transformation $(q(x),p(x))\mapsto(Q(x),P(x))$, $H\mapsto K$ be such that\footnote{At this point, 
we use 
generic notations for the canonical variables and the Hamiltonians to illustrate the main ingredients of the approach in \cite{KW,K}.}
\be
\int(P\,dQ)\,dx-K\,dt=\int(p\,dq)\,dx-H\,dt+dF\,,
\ee
for some $F$ called the generating functional of the canonical transformation. Restricting our attention to $F$ of the form
\be
F=S[q,p,Q,P]-Et
\ee
for some constant $E$,
one gets in particular,
\be
\label{transfo_Ham}
K(Q,P)=H(q,p)+E\,,
\ee
and, assuming that the new variables do not depend explicitely on $t$ and taking $q$ and $Q$ as the functionally independent variables, 
\be
\label{transfo_variables}
p=\frac{\delta S}{\delta q}~~,~~P=-\frac{\delta S}{\delta Q}\,.
\ee
For an integrable PDE, \cite{KW,K} extend \eqref{transfo_Ham} to all the integrals of the motion and require 
\be
K_n(Q,P)=H_n(q,p)+E_n\,,
\ee
or, equivalently, for the densities,
\be
{\cal K}_n(Q,P)={\cal H}_n(q,p)+\partial_x{\cal E}_n(q,p,Q,P,q_x,p_x,Q_x,P_x,\dots)\,,
\ee
in which case
\be
E_n=\left[{\cal E}_n\right]_{-\infty}^\infty\,.
\ee
Guided with this principle, they considered several well-known integrable PDEs and were able to find in each case a solution for $S$ 
such that the transformation formula \eqref{transfo_variables} yields precisely the well-known B\"acklund transformations for the model under 
consideration. 

For our purposes, the role of the two independent variables $x$ and $t$ can easily be interchanged in the previous general discussion. Therefore, 
we consider tranformations $(q(t),p(t))\mapsto(Q(t),P(t))$, $H_T\mapsto K_T$ such that 
\be
\int(P\,dQ)\,dt-K_T\,dx=\int(p\,dq)\,dt-H_T\,dx+dF_T\,,
\ee
for some $F_T$ which we regard as the generating functional of the canonical transformation with respect to the new bracket $\{~,~\}_T$. 
One also requires that all the integrals of motion (in space now) $K_{n_T}$ and $H_{n_T}$ satisfy
\be
\label{transfo_integrals}
K_{n_T}(Q,P)=H_{n_T}(q,p)+ E_{n_T}\,.
\ee
Choosing $F_T$ of the form
\be
F_T=S_T[q,p,Q,P]-E_Tx
\ee
for some constant $E_T$,
we get
\be
K_T(Q,P)=H_T(q,p)+E_T\,,
\ee
and, assuming that the new variables do not depend explicitely on $x$ and taking $q$ and $Q$ as the functionally independent variables, 
\be
\label{cano_transfo_t}
p=\frac{\delta S}{\delta q}~~,~~P=-\frac{\delta S}{\delta Q}\,.
\ee

Let us apply this approach to the situation of the sine-Gordon model in the presence of a defect described by \eqref{defect_conditions} at $x=0$. 
For $x<0$, the model is described by the fields $\tphi,\tPi$ and the integrals of motion $\widetilde{J}_n$, play the 
role of $K_{n_T}$. For $x>0$, the model is described by the field $\phi,\Pi$ and the integrals of motion $J_n$, play the 
role of $H_{n_T}$.

Therefore, in view of \eqref{transfo_gen_integrals}, condition \eqref{transfo_integrals} is easily seen to hold in our context. 
One simply expands $C(\lda)$ as $\lda\to\infty$ or $\lda\to 0$ to identify the constants $E_{n_T}$ of \eqref{transfo_integrals} from the 
defect contributions to $J_n$ and $\widetilde{J}_n$.
To complete the discussion about the canonical nature of the defect conditions, we simply have to 
find a generating functional $S_T$ such that eqs \eqref{cano_transfo_t} reproduce eqs \eqref{defect_conditions}. 
In the present situation, eqs \eqref{cano_transfo_t} take the form
\be
\label{cano_sG}
\Pi(0,t)=\frac{\delta S_T}{\delta \phi(0,t)}~~,~~\tPi(0,t)=-\frac{\delta S_T}{\delta \tphi(0,t)}\,.
\ee
Choosing
\be
\label{form_S_T}
S_T[\phi,\Pi,\tphi,\tPi]=\int_{-\infty}^\infty\cL_{defect}\,dt=\int_{-\infty}^\infty\left(\frac{1}{2}(\widetilde{\phi}\phi_t-\phi\widetilde{\phi}_t)-\cB\right)dt
\ee
where $\cB$ is given by eq \eqref{form_B}, \eqref{cano_sG} become
\be
\label{variational}
\begin{cases}
\Pi(0,t)=\frac{\partial \cL_{defect}}{\partial \phi}-\frac{\partial}{\partial t}\frac{\partial \cL_{defect}}{\partial \phi_t}\,,\\
\tPi(0,t)=-\left(\frac{\partial \cL_{defect}}{\partial \tphi}-\frac{\partial}{\partial t}\frac{\partial \cL_{defect}}{\partial \tphi_t}\right)\,.
\end{cases}
\ee
It can be verified by direct calculation that these are exactly the 
defect conditions \eqref{defect_conditions}. 
Summarizing, we have proved the following
\begin{proposition}
The defect conditions \eqref{defect_conditions} are canonical transformations for the Poisson bracket $\{~,~\}_T$. They can be conveniently written as
\be
\Pi(0,t)=\frac{\delta S_T}{\delta \phi(0,t)}~~,~~\tPi(0,t)=\frac{\delta S_T}{\delta \tphi(0,t)}\,,
\ee
where $S_T$ given in \eqref{form_S_T} and is the time integral of the defect Lagrangian density $\cL_{defect}$. The form of all the conserved quantities (in space)
is preserved in the sense of \eqref{transfo_integrals}. 
\end{proposition}
Recalling that on the solutions of the equations of motion, $\Pi=\phi_x$ and $\tPi=\tphi_x$, we note that eqs \eqref{variational} are exactly the 
defect conditions obtained in the Lagrangian approach by applying the variational principle to $\cL$ in \eqref{Lagrangian}.

\subsection{Liouville integrability: classical $r$-matrix approach with defect}

From the results of Sections \ref{Lax_new} and \ref{defect_cano}, the natural objects to study the classical $r$-matrix approach 
 for the full problem with a defect, while avoiding the usual problem, are the Poisson bracket $\{~,~\}_T$ and the following monodromy
matrix $\cM_T(x,\lda)$ defined by
\bea
\label{def_mono_matrix}
\cM_T(x,\lda)=\begin{cases} 
B_+(\lda)\widehat{\widetilde{\cT}}(x,\lda)B_-^{-1}(\lda)~~,~ x\le 0\,,\\
\widehat{\cT}(0,\lda)=B_+(\lda)\widehat{\widetilde{\cT}}(0,\lda)B_-^{-1}(\lda)~~,~~x=0\,,\\
\widehat{\cT}(x,\lda)~~,~~x\ge 0\,,  
\end{cases} 
\eea
where $B_\pm$ is given in \eqref{L_infinity}.
Note that in the traditional approach, one would consider instead $\cM_S$ in \eqref{def_M_S} to describe the system and then, one would try to compute its 
Poisson brackets with respect to $\{~,~\}_S$. The source of all complications with this approach is the evaluation of Poisson brackets involving 
$L(t,\lda)$. Instead, here, in view of Corollary \ref{Coro_Mt}, we obtain immediately, for all $x\in \RR$, 
\bea 
\label{brackets_time_mono}
\{\cM_{T1}(x,\lda),\cM_{T2}(x,\mu)\}_T=-r_+(x,\lda,\mu)\,\cM_{T1}(x,\lda)\,
\cM_{T2}(x,\mu)\nonumber\\
+\cM_{T1}(x,\lda)\,
\cM_{T2}(x,\mu)r_-(x,\lda,\mu)\,,  
\eea 
where 
\be
r_\pm(x,\lda,\mu)=\left(e^{ik_1(\lda)x\sigma_3}\otimes e^{ik_1(\mu)x\sigma_3}\right)r_\pm(\lda,\mu)\left(e^{-ik_1(\lda)x\sigma_3}\otimes e^{-ik_1(\mu)x\sigma_3}\right)
\ee
and $r_\pm$ is given by \eqref{infinite_vol_r}.
This immediately implies 
\be
\{\fa(\lda),\fa(\mu)\}_T=0~~,~~\{\widetilde{\fa}(\lda),\widetilde{\fa}(\mu)\}_T=0\,,
\ee
as desired. In particular, the integrals $J_n$ are in involution with respect 
to $\{~,~\}_T$ and the same is true of the integrals and $\widetilde{J}_n$. We can conclude that the sine-Gordon model with an integrable defect given by \eqref{defect_conditions} is 
Liouville integrable.

\section{Conclusions} \label{conclusions}

We used the idea of multisymplectic formalism to discuss Liouville integrability of the sine-Gordon model with a (B\"acklund) defect, 
thereby showing that the ideas introduced in \cite{CK}, and 
illustrated explicitely there for the nonlinear Schr\"odinger equation, apply equally well to another famous example of integrable field theory. 
The advantages of the multisymplectic formalism are threefold:
\begin{itemize}
\item  it naturally introduces a new Poisson structure for the model, in terms of which the various 
approaches to integrability in the presence of a defect given by a frozen B\"acklund transformation are unified. The bottom line 
is that a B\"acklund defect is simply a special type of canonical transformation performed at a given point in space, such that all the integrals 
of motion (in space) are preserved in form. Of course, to see this, one had first to identify the appropriate Poisson structure to be used.  
The Lagrangian approach produces 
the density appearing in the generating functional describing the defect conditions as canonical transformations with respect to the new bracket. The Lax pair 
approach produces the defect/B\"acklund matrix in terms of which the transformation formulas from the old variables to the new variables can be calculated explicitely,
as well as all the corresponding transformation formulas between old and new conserved quantities.

\item The new Poisson structure allows one to study Liouville integrability of the model with defect directly, by using a "dual" picture to the traditional one, 
whereby the system is described by a time monodromy matrix instead of the usual space monodromy matrix. When there is no defect, the two pictures are completely 
equivalent and the choice of coordinate is a matter of preference, dictated by the traditional notion of time evolution of a system. The introduction of a defect 
in space naturally distinguishes between space and time coordinates. In this case, the use of the new Poisson brackets is no longer a simple relabelling of 
coordinates but appears to be the appropriate way of treating the question of Liouville integrability in the presence of a defect.

\item The use of the new Poisson brackets shows an interesting duality in the classical $r$ matrix approach. The time part of the Lax pair satisfies the same 
Poisson algebra as the space part of the Lax pair (up to a minus sign). While this might no be so surprising for the sine-Gordon model studied in this paper
 (given the natural symmetric role of $x$ and $t$ for this model), the 
same fact was found also for the nonlinear Schr\"odinger equation in \cite{CK}. It is an interesting question to know if this holds for other well-known integrable 
systems with a Lax pair. This would suggest a connection between the multisymplectic formalism and integrable systems that goes well beyond the particular focus 
on defects that we have studied here.
\end{itemize}

Given that the method of the classical $r$ matrix goes over very naturally in our formalism, an interesting question would be to revisit the question of 
an integrable defect at the quantum level and its relation to the quantum $R$ matrix. 
For the sine-Gordon model, some aspects of this problem have already been investigated in \cite{BCZ2,CZ2} and 
quantum transmission matrices were found. Ultimately, a challenging problem would be to extend our results beyond the purely transmitting B\"acklund defects and 
find classical systems which correspond to a classical limit of the Reflection-Transmission algebras \cite{MRS,CMRS}.

\section*{Appendix} 
\appendix

\section{Proof of Eq. \eqref{equality}}\label{AppA}

We prove 
\be
\label{eq_A}
\widehat{T}_-(x,t,\lda)e^{-ik_0t\sigma_3}=\widehat{\cT}_-(x,t,\lda)e^{-ik_1x\sigma_3}\,.
\ee
Recall that, from the definition of $T(x,y,\lda)$, we obtain that $\widehat{T}(x,y,t,\lda)$ satisfies
\bea
\partial_x\,\widehat{T}(x,y,t,\lda)&=&\widehat{U}(x,t,\lda)\widehat{T}(x,y,t,\lda)\,,\\
\partial_t\,\widehat{T}(x,y,t,\lda)&=&\widehat{V}(x,t,\lda)\widehat{T}(x,y,t,\lda)-\widehat{T}(x,y,t,\lda)\widehat{V}(y,t,\lda)\,.
\eea
Let us set
\be
\Psi(x,t,\lda)=e^{i\varphi(x,t,\lda)\sigma_3}N^{-1}\widehat{T}_-(x,t,\lda)e^{-ik_0t\sigma_3}
\ee
where $\widehat{T}_-(x,t,\lda)$ is defined in \eqref{half_inf} and 
\be
\varphi(x,t,\lda)=k_1(\lda)x+k_0(\lda)t\,.
\ee
Then, one derives
\bea
\partial_x\,\Psi(x,t,\lda)&=&e^{i\varphi(x,t,\lda)\sigma_3}\cU(x,t,\lda)e^{-i\varphi(x,t,\lda)\sigma_3}\Psi(x,t,\lda)\,,\\
\partial_t\,\Psi(x,t,\lda)&=&e^{i\varphi(x,t,\lda)\sigma_3}\cV(x,t,\lda)e^{-i\varphi(x,t,\lda)\sigma_3}\Psi(x,t,\lda)\,,
\eea
where
\bea
\cU(x,t,\lda)&=&N^{-1}(\widehat{U}(x,t,\lda)-U_\infty)N\,,\\
\cV(x,t,\lda)&=&N^{-1}(\widehat{V}(x,t,\lda)-V_\infty)N\,.
\eea
Consequently, note that, 
\be
\lim_{x\to-\infty}\Psi(x,t,\lda)=\1\,,
\ee
and therefore we can write
\be
\Psi(x,t,\lda)=\1+\int_{-\infty}^xe^{i\varphi(\xi,t,\lda)\sigma_3}\cU(\xi,t,\lda)e^{-i\varphi(\xi,t,\lda)\sigma_3}\Psi(\xi,t,\lda)\,d\xi\,.
\ee
Given that 
\be
\lim_{t\to -\infty}\cU(x,t,\lda)=0\,,
\ee
we deduce that 
\be
\lim_{t\to -\infty}\Psi(x,t,\lda)=\1\,.
\ee
Similarly, from the definition of $\cT(t,\tau,\lda)$ and $\widehat{\cT}_-(t,x,\lda)$ in \eqref{half_inf}, if we set 
\be
\Phi(x,t,\lda)=e^{i\varphi(x,t,\lda)\sigma_3}N^{-1}\widehat{\cT}_-(x,t,\lda)e^{-ik_1x\sigma_3}\,,
\ee
we find that 
\bea
\partial_x\,\Phi(x,t,\lda)&=&e^{i\varphi(x,t,\lda)\sigma_3}\cU(x,t,\lda)e^{-i\varphi(x,t,\lda)\sigma_3}\Phi(x,t,\lda)\,,\\
\partial_t\,\Phi(x,t,\lda)&=&e^{i\varphi(x,t,\lda)\sigma_3}\cV(x,t,\lda)e^{-i\varphi(x,t,\lda)\sigma_3}\Phi(x,t,\lda)\,,
\eea
and 
\be
\lim_{t\to -\infty}\Phi(x,t,\lda)=\1\,.
\ee
So we can write
\be
\Phi(x,t,\lda)=\1+\int_{-\infty}^te^{i\varphi(x,\tau,\lda)\sigma_3}\cV(x,\tau,\lda)e^{-i\varphi(x,\tau,\lda)\sigma_3}\Phi(x,\tau,\lda)\,d\tau\,,
\ee
and then deduce
\be
\lim_{x\to -\infty}\Phi(x,t,\lda)=\1\,.
\ee
Hence, both $\Psi$ and $\Phi$ satisfy the same differential equations with same boundary conditions: they must be equal, which is precisely \eqref{eq_A}.
Of course, a similar statement can be made for $\widehat{T}_+(x,t,\lda)$ and $\widehat{\cT}_+(x,t,\lda)$.

\end{document}